# High performance bilayer-graphene Terahertz detectors

Davide Spirito,[1] Dominique Coquillat,[2] Sergio L. De Bonis,[1] Antonio Lombardo,[3] Matteo Bruna[3], Andrea C. Ferrari,[3] Vittorio Pellegrini,[4,1] Alessandro Tredicucci,[1] Wojciech Knap,[2] and Miriam S. Vitiello[1*]

[1] *NEST, Istituto Nanoscienze – CNR and Scuola Normale Superiore, Piazza San Silvestro 12, Pisa, I-56127*
[2] *Laboratoire Charles Coulomb UMR 5221 Université Montpellier 2 & CNRS-, F-34095, Montpellier, France*
[3] *Cambridge Graphene Centre, Cambridge University, Cambridge, CB3 0FA, UK*
[4] *Istituto Italiano di Tecnologia (IIT), Via Morego 30, 16163 Genova, Italy*

*Corresponding author: miriam.vitiello@sns.it*

**Abstract** We report bilayer-graphene field effect transistors operating as THz broadband photodetectors based on plasma-waves excitation. By employing wide-gate geometries or buried gate configurations, we achieve a responsivity~1.2V/W (1.3 mA/W) and a noise equivalent power~$2\times10^{-9}$ W/√Hz in the 0.29-0.38 THz range, in photovoltage and photocurrent mode. The potential of this technology for scalability to higher frequencies and the development of flexible devices makes our approach competitive for a future generation of THz detection systems.

Generation and detection of radiation across the far infrared or Terahertz (THz) region of the electromagnetic spectrum is promising for a large variety of strategic applications, [1] ranging from biomedical diagnostics [2] to process and quality control,[3] homeland security[1] and environmental monitoring.[4] Due to its non-ionizing nature,[1] THz radiation can penetrate many commonly used dielectrics,[1] otherwise opaque for visible and mid-infrared light, allowing detection of specific spectroscopic features [5] with a sub-millimeter diffraction-limited resolution.[1]

Quantum cascade (QC) lasers[6] are nowadays capable to reach more than 120mW output powers in continuous wave, and to operate in a broad frequency range (1.2 - 4.7 THz). [7] However, a low-cost room-temperature (RT) THz detector technology with hundreds V/W responsivity, and fast (GHz modulation frequency) is still missing. Conventional THz and sub-THz detection systems based on incoherent (pyroelectric, Golay cell, Si bolometers)[8] or coherent (heterodyne mixers)[8] approaches are either very slow (~ 100 Hz modulation frequency)[8] or require deep cryogenic cooling,[8] or, if exploiting fast (hundred MHz) nonlinear electronics, show a significant drop of performance above 1 THz.[8] Recently,[9-13] detectors based on the gate-modulation of the channel conductivity by the incoming radiation have been realized, employing III-V materials in high-electron-mobility transistors (HEMT), [9,10] two dimensional electron gas field effect transistors (FET),[9] Si- metal–oxide–semiconductor FETs and complementary metal-oxide semiconductor architectures,[11] showing fast response times and high responsivities. This approach was also extended[12] to InAs nanowire one dimensional FETs, allowing RT detection of THz emission in the 0.3 – 3 THz range.[12,13]

The operating mechanism of a FET detector [14] can intuitively be interpreted as deriving from the nonlinearity arising from the simultaneous modulation of carrier density and drift velocity by the THz radiation.[14] When a THz beam is funneled onto the channel of a FET, it modulates the gate and source/drain potentials, generating a continuous source-drain voltage $\Delta u$ (or current $\Delta i$)[14] which can be maximized by varying the gate bias $V_G$. This modulation leads to excitation of plasma oscillations that can either propagate in the channel (plasma waves)[14] or be damped (overdamped plasma waves).[14] At THz frequencies plasma waves decay on a distance $d_w = s\tau$, [14] where $s$ is the plasma wave velocity (typically assumed $\sim 10^8$ cm/s independently from the channel material) [14] and $\tau$ the momentum relaxation time. If the FET gate length $w_G > d_w$, plasma waves excited at the source decay before reaching the drain, and broadband THz detection is achieved.[9] The same condition occurs in the low frequency regime $2\pi\upsilon\tau <1$ ($\upsilon$ being the frequency of the THz radiation), where the plasma oscillations decay with a frequency-dependent characteristic length [9,14] $l = s\left(\frac{2\tau}{\omega}\right)^{1/2}$.

The unique optoelectronic properties of graphene make it an ideal platform for a variety of photonic applications. [15,16] In particular, due to its high carrier mobility, gapless spectrum, and frequency-independent absorption, [16-18] graphene is very promising material for the development of THz detectors,[19] still severely lacking in terms of solid-state devices.

Refs. 20, 21 theoretically suggested that THz detection in graphene could be mediated by electron-hole pairs or interband photo-generation. However, we experimentally reported RT broadband THz detection as a result of overdamped plasma-waves excitation in the channel of a graphene-FET[19] operating in photovoltage mode, with an electrical output per optical input (responsivity) $R_v \sim$ 0.015 V/W and noise equivalent powers NEP = 30 nWHz$^{1/2}$. Subsequently, Ref. 22 reported ultrafast (10 ps response time) THz detectors based on photo-induced bolometric effects in graphene, operating in photocurrent mode, with a responsivity ~ 5 nA/W. Ref. 23 then combined plasmonic/bolometric graphene FETs, reaching $R_v \sim$ 150 µV/W and operating in photovoltage mode at much higher frequencies (2.5 THz).

Here we report THz detectors operating at RT in either photovoltage or photocurrent mode with a significant enhancement of sensitivity and lowered NEPs compared to the state of the art. This is achieved by using either large (1 µm) gate lengths, or buried gate geometries on a bilayer graphene (BLG) FET. We use BLG instead of single layer graphene (SLG) since the modulation of carrier density was proved to be more effective in the former case, [19] allowing a higher responsivity at THz frequencies.

The devices are prepared as follows. Flakes are mechanically exfoliated from graphite [24] on an intrinsic Si substrate covered with 300nm SiO$_2$. BLGs are selected and identified by a combination of optical microscopy[25] and Raman spectroscopy. [26,27] These are then used to fabricate two sets of FETs (samples A, B). In A, source (*S*) and drain (*D*) contacts are patterned by electron beam lithography (EBL) and metal evaporation (5nm Cr /80 nm Au). The *S* contact is connected to one lobe of a 50° bow-tie antenna, and *D* to a metal pad. After placement of 35nm HfO$_2$ by atomic layer deposition (ALD), the other lobe of the antenna is fabricated, and constitutes the FET gate (*g*). The channel length is $L_{SD}$ = 2.5 µm, while the gate length $w_G$ =1 µm (Fig. 1a). In B, one lobe of a log-periodic circular-toothed antenna is fabricated by EBL and metal evaporation to act as buried gate. After deposition of 35 nm HfO$_2$ by ALD, *S* and *D* electrodes are fabricated. Similar to sample A, *S* is the second lobe of the antenna, while the drain is a metal line connecting to a bonding pad. The channel length is 2.5 µm, its width 7.5 µm, while $w_G$ = 0.3µm (Fig. 1b). A BLG flake is then placed onto the pre-fabricated electrodes by wet transfer.[28] Fig. 2 compares the Raman spectra of the flake prior and after placement onto the electrodes. The "2D" peak shows the characteristic multi-band shape of BLGs.[26,27] The absence of a significant Raman "D" peak prior and after transfer indicates the low-defect samples are not damaged by this process. Scanning electron microscope images of both devices are reported in Fig. 1c,d.

The devices are electrically characterized by means of two voltage generators to drive independently $V_G$ and the source-to-drain ($V_{SD}$) voltage. The drain contact is connected to a current amplifier converting the current into a voltage signal with an amplification factor of $10^4$ V/A. The latter signal is then measured with an Agilent 34401A voltmeter reader. The current ($I_{SD}$) as a function of $V_G$ characteristics are measured while sweeping $V_G$, keeping $V_{SD}$ = 0.001V.

Two sets of experiments are performed in order to compare the FET transport behavior: one under 0.292 THz irradiation, the other without. Figs 3a-b show that the conductivity, σ goes through a minimum when the chemical potential below the gate crosses the charge neutrality point (*CNP*), as in Ref. 19. Moreover, a shift of the *CNP* of ~ 0.1V is detected when the THz beam is focused on the sample. The channel resistance (R) varies from 2.8 to 4.2kΩ in A, and from 500 to 800Ω in B, corresponding to a mobility µ ≈ 3000 cm$^2$/Vs and a carrier density n$_o$ ≈ 2×10$^{11}$cm$^{-2}$, and  µ ≈ 200 cm$^2$/Vs, n$_o$ ≈ 70 ×10$^{11}$cm$^{-2}$, respectively, as extracted from the transfer characteristics, following the procedure described in Ref. 29. This implies a plasma wave decay distance $d_w$ ≈ 1.5µm ≥ $w_G$ (A) and *l* ≈ 350 nm ≥ $w_G$ (B), meaning that both are overdamped.

According to a diffusive model of transport,[14,30,31] for long gates ($w_G$ >> $d_w$) or in a low-frequency regime (2πυτ <1), a second-order nonlinear response is expected in a FET when an oscillating THz is applied between *G* and *S*. In this case the photovoltage signal can be qualitatively extracted from the current-voltage characteristics via the equation [14] , $\Delta u = K\, \sigma^{-1} \frac{d\sigma}{dV_G}$ where the constant *K* represents the coupling efficiency of the incoming radiation to the antenna together with a loading factor determined by the impedance of the transistor. The $\sigma^{-1} \frac{d\sigma}{dV_G}$ trend shown on the left vertical axis of Figs. 2c,d indicates a clear D*u* variation from negative to positive values around the CNP for A, consistent with ambipolar transport,[28] while only a small sign switch is recorder for B, in agreement with the evident asymmetry of the transfer characteristics.[18] On the right vertical axis of Figs 3c-d we plot the photovoltage value $\Delta u^* = (I_{SD,THz} - I_{SD,no\ THz})\sigma^{-1}$ extracted from the difference between the current measured with and without the THz beam. $\Delta u^*$ well reproduces the expected THz photovoltage signal.

To experimentally access the photoresponse directly, we employ both the 0.292 THz radiation generated by a fixed frequency (Gunn diode) electronic source and a frequency tunable electronic source covering the 0.265-0.375 THz range. The THz beam is collimated and focused by a set of two off-axis parabolic mirrors, while the intensity is mechanically chopped at a frequency of 473 Hz. Two sets of experiments are performed: i) the photovoltage signal is recorded by means of a lock-in amplifier in series with a voltage preamplifier, having an input impedance of 10 MΩ and an amplification factor $G_n = 25$; ii) the photocurrent signal is measured with a transimpedance amplifier, having amplification factor $G_n = 10^7$ V/A. In all cases, the vertically polarized incoming radiation impinges on the FET mounted in a dual-in-line package, with an optical power $P_t = 1.8$ mW (fixed frequency source) or $P_t = 0.34$ mW (tunable source).

Figure 4a plots $R_v$ extracted from the photovoltage measurements on A, while focusing the 0.292 THz radiation on it and keeping $V_{SD}=0$. $R_v$ is related to the measured photo-induced voltage $\Delta u(V_G) = \frac{2\sqrt{2}\frac{\pi}{4} LIA}{G_n}$ via : $R_v = \frac{\Delta u S_t}{S_a P_t}$, where $S_t$ is the radiation beam spot area, $S_a$ is the detector active area and $LIA$ the lock-in signal.[19] Loading effects due to the finite impedance of the preamplifier input are neglected. This procedure assumes that the entire power incident on the antenna is fully coupled to the FET, meaning that the present $R_v$ is a lower limit. Taking into account our beam spot diameter $d \approx 4$ mm, we have $S_t = \pi d^2/4 = 12.6 \times 10^{-6}$ m$^2$. Since the total detector area is smaller than the diffraction limited one $S_\lambda = \lambda^2/4$, the active area is taken equal to $S_\lambda$. The dependence of $\Delta u$ on $V_G$ is in qualitative agreement with the trend shown in Fig. 3c, thereby proving that the detector still operates in the broadband overdamped regime. Note that the sign of the photovoltage changes abruptly in the vicinity of the CNP, following the switch of sign of the derivative in $\sigma^{-1}\frac{d\sigma}{dV_G}$. For a qualitative evaluation of the coupling efficiency of the THz radiation by the antenna, we can use the outcome of a fitting procedure to the responsivity of other FET detectors with similar antenna designs measured in the same experimental set-up,[31] giving $\square = 1.25\times10^{-4}$ V$^2$. By using the extrapolated $K$ value, the predicted photovoltage (Fig. 3c) is in excellent agreement with Fig. 3a. The comparison with $\Delta u^*$ shows a good agreement with the measured $R_v$, with a small voltage offset possibly due to loading impedance and/or heating effects. We get $\Delta u/\Delta u^* \sim 0.25$, as expected if one considers that $\Delta I \times \sigma^{-1}$ is much less influenced by the loading than the photovoltage, since the former is a purely *dc* measurement, not affected by the device capacitive reactance.

Maximum responsivities of 1.2 V/W, i.e. more than one order of magnitude larger than in previous reports[10,19,23] are reached. This is interpreted as due to the larger gate length, as predicted by the photo-induced voltage equation:[14] $\Delta u(x) \propto \left[1 - \exp\left(\frac{-2x}{l}\right)\right]$ where $x$ is the gated region length and $l$ the previously defined characteristic decay length of the *ac* voltage (for sample A is $l \sim 1.2$ μm).[19] The ratio $w_G/l \sim 0.8$ is then more than one order of magnitude larger than in previous BLG-FET THz detectors ($w_G/l = 0.03$),[19] meaning that any competing thermoelectric effects possibly arising from the ungated regions[19] are here less relevant. Also, no evidence of THz driven interband transitions[19] is seen.

Figure 4b plots $R_v$ extracted from the photovoltage measurements on B at both 0.292 and 0.358 THz. The $R_v$ vs $V_G$ trend is almost independent from the radiation frequency (confirming the broadband detection mechanism) with maximum $R_v$ = 1.12 V/W. For comparison, Fig. 4c plots the responsivity values obtained while B is operating in photocurrent mode at 0.358 THz. The maximum $R_v$ = 1.3 mA/W is in agreement with that recorded while the device is operating in the photovoltaic mode, once the 800Ω channel resistance is taken into account. The $w_G/l$ ratio is here ≈ 0.8, i.e. comparable to A, motivating the equivalent maximum responsivity values. Also in agreement with A, we have $Δu^* > Δu$ (see Fig. 3d), being $Δu/Δu^* \sim 0.5$.

Figure 5a plots the photovoltage value measured at $V_G$ = 2.5V, while varying the frequency of the incoming THz beam. The frequency response has several peaks, as expected from the broadband nature of the antenna, with the broader and more intense band at 0.358 THz.

Figure 5b shows the Johnson–Nyquist noise extracted from the data of Fig. 3b, when B is operating in photovoltage mode $N_V = \sqrt{4kTR}$, or photocurrent mode $N_I = \sqrt{\frac{4kT}{R}}$. Assuming that the Johnson–Nyquist noise is the dominant contribution,[32] as typical for THz FET detectors,[12,13,19] minimum $NEP \approx 4\times10^{-9}$ W/Hz$^{1/2}$ and $NEP \approx 2\times10^{-9}$ W/Hz$^{1/2}$ are reached in photovoltage and photocurrent mode, respectively, comparable with most commercially available RT THz detectors.[8]

In summary, by employing wide-gate geometries or buried gate configurations, we fabricated bilayer-graphene FETs with $R_v \approx$ 1.2V/W (1.3 mA/W) and NEP ≈ $2\times10^{-9}$ W/√Hz. These value are competitive with the performance of commercially available detection systems, paving the way to a realistic exploitation for large-area, fast imaging of macroscopic samples, spectroscopic applications, process control and biomedical diagnostics.

**Acknowledgments**

We acknowledge funding from the Italian Ministry of Education, University, and Research (MIUR) through the program "FIRB - Futuro in Ricerca 2010" RBFR10LULP "Fundamental research on Terahertz photonic devices" and RBFR10M5BT PLASMOGRAPH by EU Graphene Flagship (contract no. CNECT-ICT-604391), ERC projects NANOPOTS, Hetero2D, EU projects GENIUS, CARERAMM, MEM4WIN, a Royal Society Wolfson Research Merit Award, EPSRC grants EP/K01711X/1, EP/K017144/1, the Italian Ministry of Economic Development through the ICE-CRUI project "Teragraph", the CNRS through the GDR-I project "Semiconductor sources and detectors of THz frequencies", the Region Languedoc-Roussillon Terahertz Platform.

**Figure captions**

**Figure 1:** (a-b) Schematic BLG-FETs with (a) top gate, (b) buried gate. (c-d) False-color SEM micrographs of the devices schematically shown in (c) panel a, and (d), panel b.

**Figure 2:** Measured Raman spectra of the BLG flake prior (bottom curves) and after (upper curves) placement onto the electrodes.

**Figure 3:** (a-b) Source-drain current ($I_{SD}$) measured as a function of $V_G$ in (a) sample A and (b) B, while keeping $V_{SD}$ = 1mV and with (red curve) or without (black curve) a 0.292 THz illumination. (c-d) Left vertical axis: Derivative of the conductivity multiplied by the resistance, as a function of $V_G$, for (c) sample A, and (d) sample B. Right vertical axis: difference between the current measured while focusing the THz beam on the sample $I_{SD,THz}$ and without it $I_{SD,no\ THz}$, multiplied by the channel resistance in case of (c) sample A, or (d) sample B. The dashed vertical lines indicate the charge neutrality point.

**Figure 4:** (a) Room-temperature (RT) responsivity as a function of $V_G$ measured while sample A is operating in photovoltage mode and while keeping the polarization of the incoming 0.292 THz beam parallel to the antenna axis. The related measured photovoltage signal (blue curve) is shown on the right vertical axis. (b) RT responsivity as a function of $V_G$ measured while sample B is operating in photovoltage mode and while keeping the polarization of the incoming 0.292 THz (upper red curve) or 0.358 THz (pink curve) beams parallel to the antenna axis. The related measured photovoltage signals (blue and black curves) are shown on the right vertical axis. c) RT responsivity as a function of $V_G$ measured while sample B is operating in photocurrent mode (c) at 0.358 THz. The dashed line marks the zero value of output voltage.

**Figure 5:** (a) Photocurrent ($\Delta i$) and photovoltage ($\Delta u$) signals measured in sample B as a function of frequency of the incoming THz beam while $V_G$ = 2.4 V and $V_{SD}$= 1 mV. (b) Johnson–Nyquist noise extracted from the data of Fig. 2b for sample B while it is operating in photocurrent ($N_I$) or photovoltage mode ($N_V$).

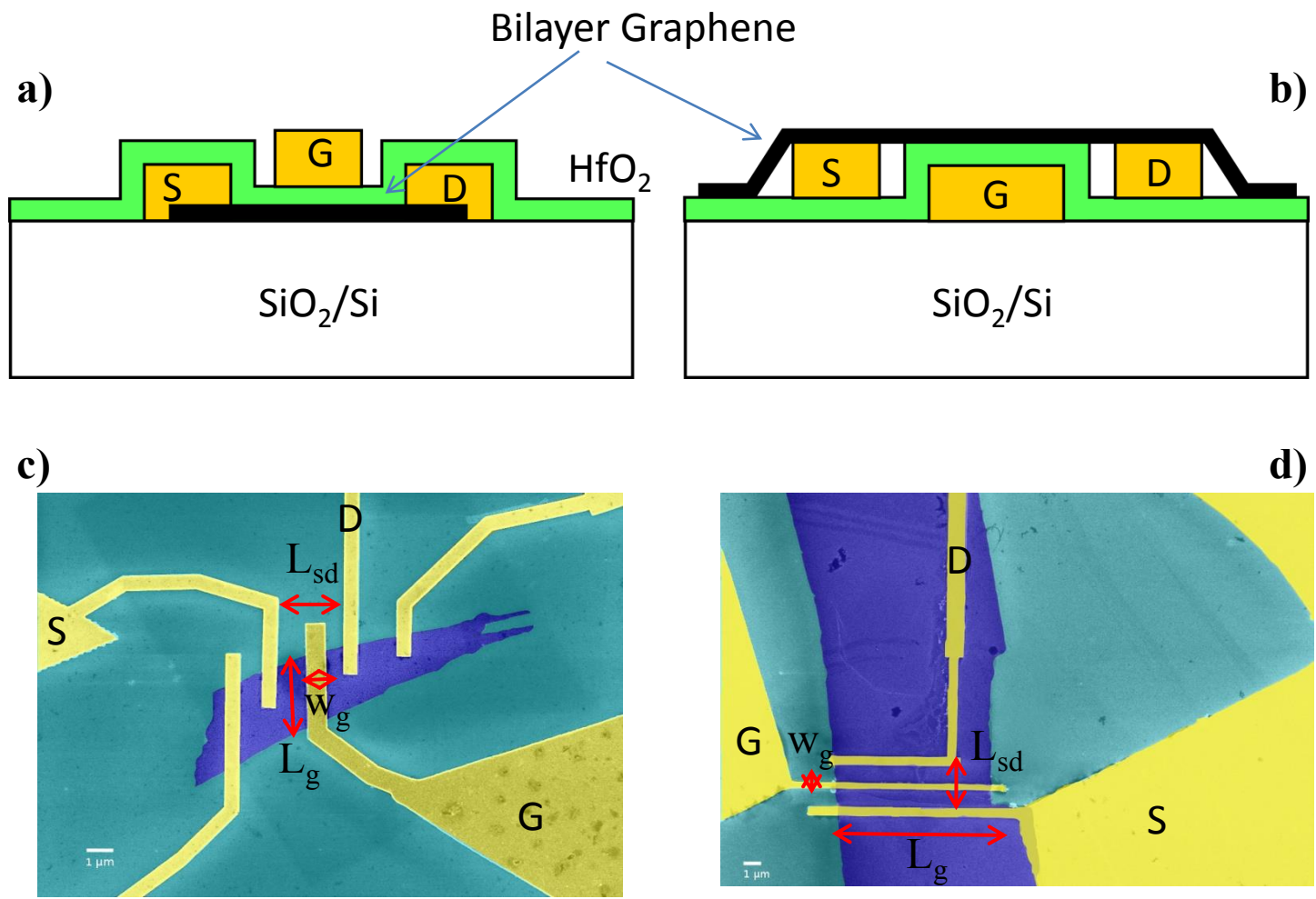

**Figure 1**

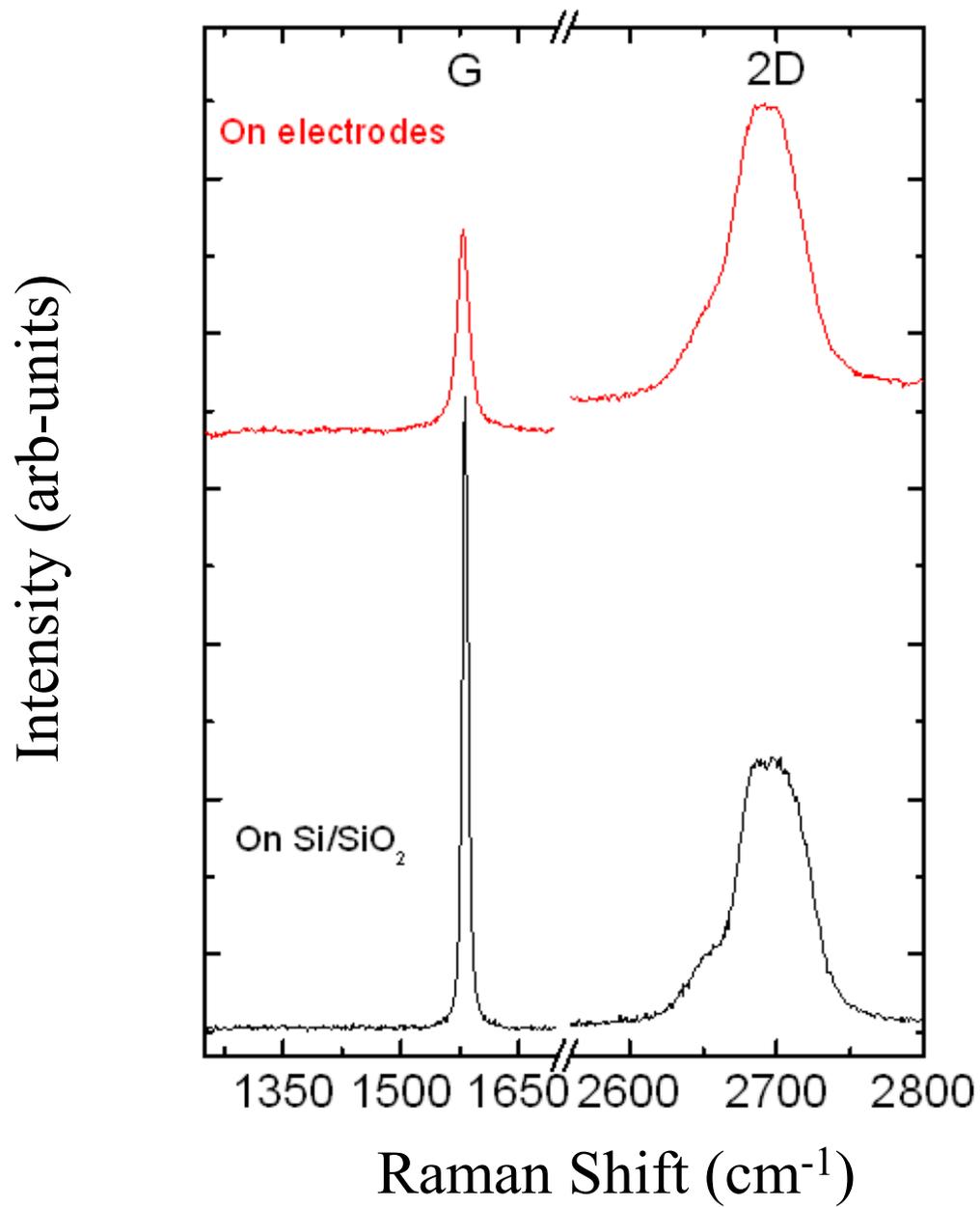

**Figure 2**

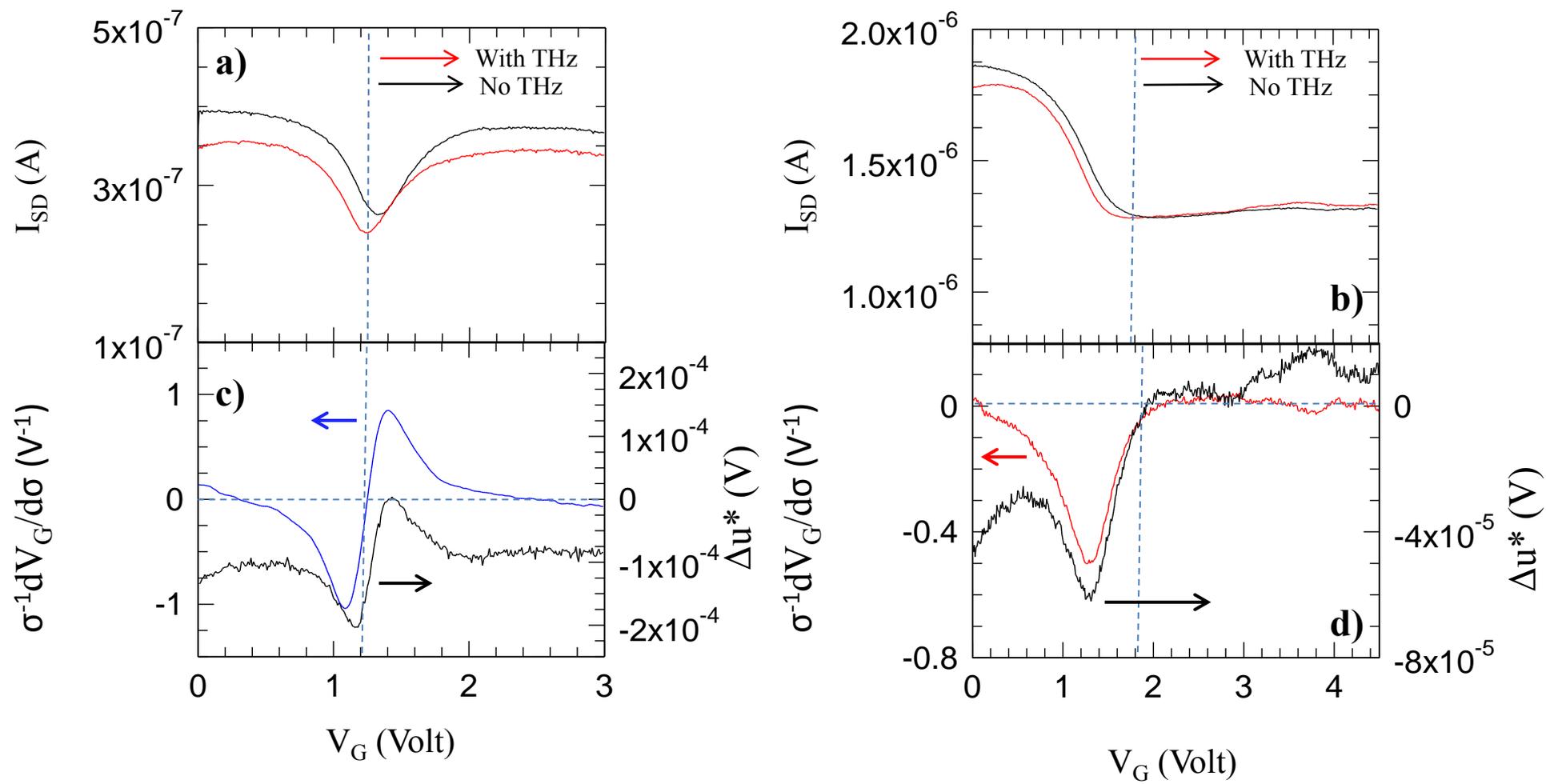

**Figure 3**

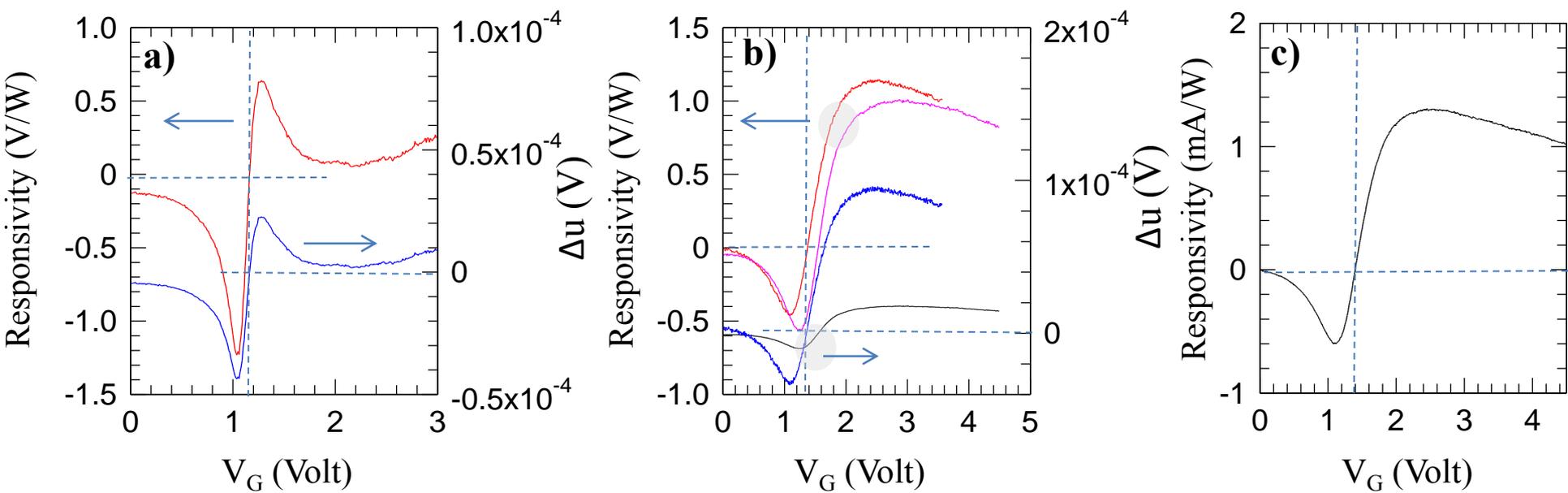

**Figure 4**

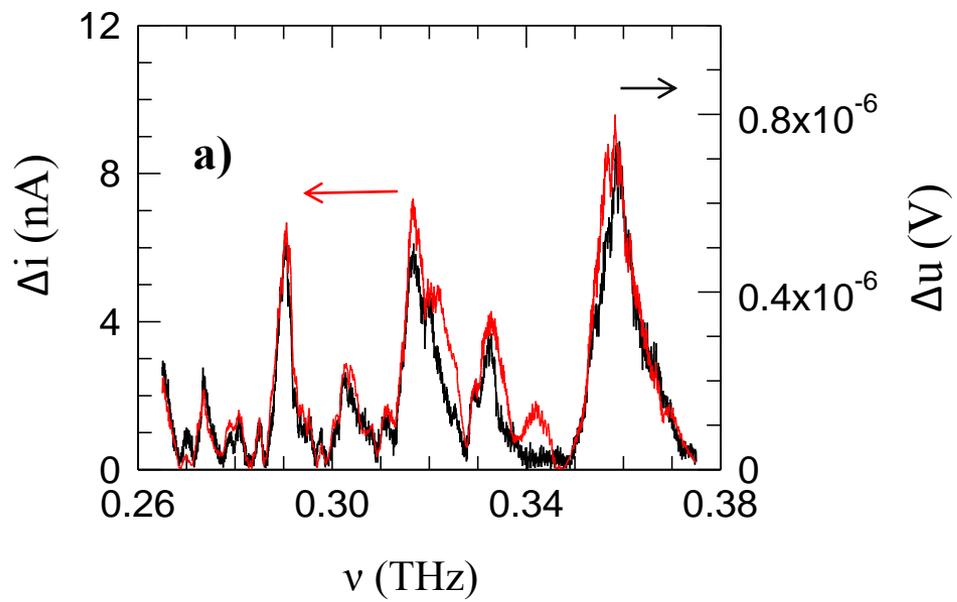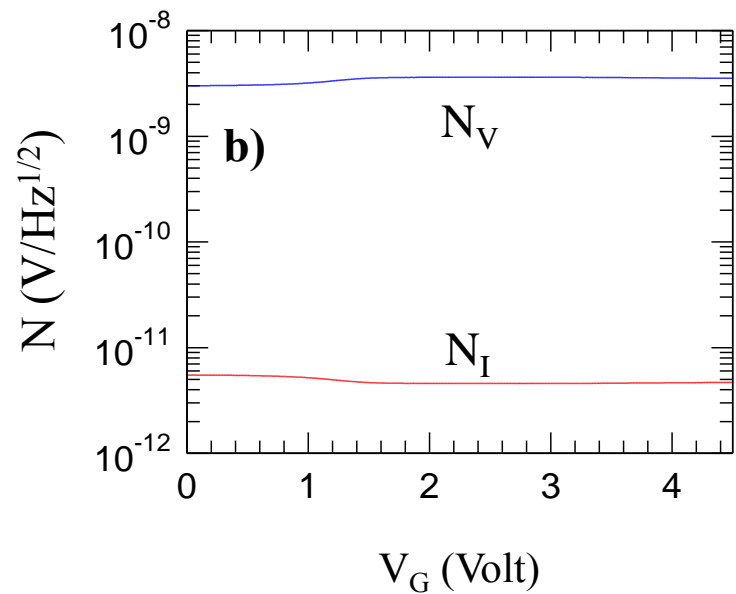

**Figure 5**